# A Deep Learning Approach to Interface Color Quality Assessment in HCI

Shixiao Wang*
School of Visual Arts
New York, USA

Runsheng Zhang
University of Southern California
Los Angeles, USA

Junliang Du
Shanghai Jiao Tong University
Shanghai, China

Ran Hao
University of North Carolina at Chapel Hill
Chapel Hill, USA

Jiacheng Hu
Tulane University
New Orleans, USA

*Abstract*-**In this paper, a quantitative evaluation model for the color quality of human-computer interaction interfaces is proposed by combining deep convolutional neural networks (CNN). By extracting multidimensional features of interface images, including hue, brightness, purity, etc., CNN is used for efficient feature modeling and quantitative analysis, and the relationship between interface design and user perception is studied. The experiment is based on multiple international mainstream website interface datasets, covering e-commerce platforms, social media, education platforms, etc., and verifies the evaluation effect of the model on indicators such as contrast, clarity, color coordination, and visual appeal. The results show that the CNN evaluation is highly consistent with the user rating, with a correlation coefficient of up to 0.96, and it also shows high accuracy in mean square error and absolute error. Compared with traditional experience-based evaluation methods, the proposed model can efficiently and scientifically capture the visual characteristics of the interface and avoid the influence of subjective factors. Future research can explore the introduction of multimodal data (such as text and interactive behavior) into the model to further enhance the evaluation ability of dynamic interfaces and expand it to fields such as smart homes, medical systems, and virtual reality. This paper provides new methods and new ideas for the scientific evaluation and optimization of interface design.**

*Keywords-Deep convolutional neural network; interface color quality; quantitative evaluation; human-computer interaction*

I. INTRODUCTION

With the rapid development of information technology, human-computer interaction interfaces play an increasingly important role in people's daily lives and work. From mobile applications to industrial control systems, human-computer interaction interfaces not only assume the function of information transmission but are also the key carriers of user experience [1]. The design quality of the interface directly affects the user's operating efficiency, experience satisfaction, and the market competitiveness of the product. Among many design factors, color, as the primary element of visual communication, has an important impact on the user's perception and emotion [2]. However, traditional interface design and evaluation mostly rely on the designer's experience and subjective judgment, lacking scientific and systematic quantitative analysis methods, resulting in the difficulty of quantifying the evaluation of design schemes and limited optimization space [3]. Therefore, exploring the interface color quality evaluation model based on scientific methods has important research value and application prospects.

In recent years, deep learning technology, especially deep convolutional neural networks (CNN), has shown excellent performance in the field of computer vision [4-5]and has been widely used in image classification [6], target detection [7], image generation [8], and other tasks. Compared with traditional methods, CNN can automatically extract multi-level features of images, from low-level edge information to high-level semantic expression, providing a powerful tool for solving complex image analysis problems. The color characteristics and aesthetic quality evaluation of human-computer interaction interfaces involve multi-dimensional features such as image color, texture, and layout, which can be efficiently extracted and modeled by deep convolutional networks. Therefore, introducing deep learning into the quantitative study of interface color quality provides a new methodology for the optimization of human-computer interaction interfaces [9].

This study aims to combine deep convolutional networks to build a set of evaluation models suitable for the quantification of color quality of human-computer interaction interfaces. By analyzing the color distribution, hue contrast, brightness level, and purity coordination of interface images, combined with high-dimensional features extracted by deep convolutional networks, the impact of interface design on user aesthetics is quantified [10]. The study not only involves the color quality evaluation of a single interface element but also includes the balance, unity, and coherence analysis of the overall interface layout. With the help of deep learning models, the experiment will further explore the implicit relationship between color features and user preferences and provide data-driven theoretical support for the optimization of interface design.

## II. RELATED WORK

Deep learning has significantly advanced the evaluation of human-computer interaction (HCI) interfaces by enabling automated and data-driven analysis of design quality. A critical aspect of interface design is understanding how users perceive visual elements. Duan [11] systematically analyzed user perception in interface design and emphasized the importance of data-driven evaluation methods. This aligns with the objective of this study, which seeks to provide a structured and reproducible framework for quantifying interface color quality. Sun et al. [12] explored adaptive user interface generation using reinforcement learning, demonstrating the potential of optimizing interface layouts based on extracted feature representations, a concept relevant to automated interface evaluation. Duan [13] and Shao et al. [14] investigated deep learning techniques for gesture recognition, which rely on effective feature extraction to understand user interactions—an approach that parallels the use of CNNs in interface quality assessment by capturing complex spatial and structural relationships within images.

Feature extraction remains a crucial challenge in deep learning-based evaluation models. CNNs have demonstrated strong capability in identifying structural patterns, making them highly suitable for interface analysis. Du [15] introduced an optimized CNN model that improves feature extraction efficiency, reinforcing the idea that well-structured convolutional architectures can enhance automated evaluation tasks. Wang [16] explored multimodal data mining using sparse decomposition and adaptive weighting, an approach that aligns with this study's objective of capturing and analyzing multidimensional visual properties, such as hue, brightness, and contrast. Similarly, Gao et al. [17] proposed a multi-level attention mechanism to refine feature representations, which could further enhance the accuracy of interface color quality assessment by allowing the model to focus on subtle yet important design characteristics.

Advancements in transformer-based architectures have contributed to improved feature alignment and classification performance in deep learning applications. Wang [18] applied transformers and graph neural networks (GNNs) to analyze sequential patterns, highlighting the effectiveness of structured learning in complex data environments, a methodology that could be extended to track visual patterns across multiple interface samples. Scalability and computational efficiency are crucial for large-scale interface evaluations. Sun [19] developed a dynamic distributed scheduling approach to optimize computational load, offering valuable insights into the efficient processing of large-scale datasets. While this study primarily focuses on CNN-based evaluation, scalability remains an important consideration, particularly for real-time interface assessments and dynamic UI adaptations. Li [20] introduced an improved transformer model for cross-domain feature alignment, demonstrating the potential of refining feature representations, which could be leveraged to enhance the robustness of interface quality evaluation. These studies provide essential insights into improving deep learning models by refining feature relationships, thereby strengthening the analytical depth of CNN-based assessments.

Building upon these prior works, this study integrates CNN-based feature extraction techniques with quantitative evaluation metrics to assess interface color quality objectively. Unlike heuristic or purely subjective approaches, this model systematically quantifies visual aesthetics using deep learning, making interface evaluation more consistent, scalable, and reproducible. The incorporation of hierarchical feature extraction, attention mechanisms, and adaptive processing techniques provides a robust foundation for improving interface design evaluation, contributing to the broader goal of intelligent and data-driven interface optimization.

## III. METHOD

In this study, we combined deep convolutional neural networks (CNNs) to build a color quality quantification model for human-computer interaction interfaces [21]. Specifically, the core of the method is to extract the color features of the human-computer interaction interface through a deep convolutional network and build a quantitative evaluation model based on these features. Its network architecture is shown in Figure 1.

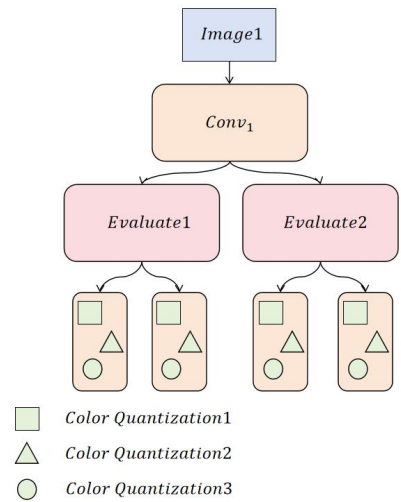

Figure 1 Overall model architecture

First, for the input human-computer interaction interface image, we represent it as a three-dimensional tensor $I \in R^{H \times W \times C}$ where H, W, and C represent the height, width, and number of channels of the image (usually C=3, corresponding to RGB three channels). In order to ensure the compatibility of the model with images of different resolutions, we uniformly scale the input image to a fixed size $H_0 \times W_0$. Next, the image $I$ is extracted through a deep convolutional neural network to extract a multi-layer feature map $F_l$, that is:

$$F_l = Conv_l(F_{l-1}), \quad F_0 = I$$

Where $Conv_l(\cdot)$ represents the convolution operation of the l-th layer, $F_l \in R^{H_l \times W_l \times D_l}$ is the feature map of the l-th layer, and $D_l$ represents the number of channels of this layer.

The extracted feature map $F_l$ contains multi-level information of the image. In this study, we focus on the following color-related features: hue distribution, brightness distribution, and purity distribution. We perform global pooling on the feature map $F_L$ (the deepest feature) to obtain high-level semantic information, while performing local statistical analysis on the shallow feature $F_1, F_2, ..., F_k$ to retain fine-grained color information. The final feature representation is:

$$z = Concat(GAP(F_L), Stat(F_1, F_2, ..., F_k))$$

$GAP(\cdot)$ represents global average pooling, $Stat(\cdot)$ represents the calculation of statistical features such as mean and standard deviation of shallow features, and $Concat(\cdot)$ represents feature concatenation.

Next, we define the evaluation model of interface color quality. Based on the aesthetic evaluation research in psychology, the evaluation of interface quality can be expressed as a weighted sum model:

$$Q = \alpha \cdot Q_{hue} + \beta \cdot Q_{lightness} + \gamma \cdot Q_{purity}$$

Where $Q_{hue}$, $Q_{lightness}$ and $Q_{purity}$ represent the quality scores of hue, lightness and purity respectively, and $\alpha$, $\beta$ and $\gamma$ are the corresponding weight coefficients, satisfying $\alpha + \beta + \gamma = 1$.

We further construct a calculation method for each quality score [22]. Taking hue quality as an example, we calculate the hue distribution $Hue(I)$ of the interface image in the HSV color space and calculate the entropy value of the hue to measure the uniformity of its distribution:

$$Q_{hue} = -\sum_{i=1}^{n} p_i \log p_i$$

Where $p_i$ represents the probability of the hue value being in the i-th interval, and $n$ represents the total number of intervals. Similarly, the quality scores for brightness and purity can be defined by a combination of distribution uniformity and contrast, specifically:

$$Q_{lightness} = \frac{\sigma_{lightness}}{\mu_{lightness}}, Q_{purity} = \frac{\sigma_{purity}}{\mu_{purity}}$$

$\sigma_{lightness}$ and $\sigma_{purity}$ are the standard deviations of brightness and purity respectively, and $\mu_{lightness}$ and $\mu_{purity}$ are their means.

Finally, we use supervised learning to train the evaluation model. Given an interface image dataset $\{l_i, y_i\}_{i=1}^{N}$, where $y_i$ is the manually annotated quality score of the i-th interface, the loss function is defined as:

$$L = \frac{1}{N} \sum_{i=1}^{N} (Q(l_i; \theta) - y_i)^2$$

Where $Q(l_i; \theta)$ represents the model's predicted quality score for the i-th interface, and $\theta$ represents the model's parameters. By optimizing $L$, we learn the optimal model parameters.

The above method provides a complete color quality quantification process, from feature extraction to quantification calculation, to model training and optimization, to ensure that the model can effectively evaluate the color quality of the human-computer interaction interface.

IV. EXPERIMENT

A. Datasets

The dataset used in this study comes from a variety of website pages publicly available on the Internet, covering a wide range of fields and application scenarios. First, the dataset includes the homepages of many internationally renowned shopping platforms, such as Amazon, eBay, and Walmart. These pages provide diverse color features for research due to their rich color matching and complex interface design. The pages of such e-commerce platforms usually use high-contrast color schemes to attract users' attention, and contain a large number of pictures, texts, and dynamic elements, making them an ideal data source for analyzing color quality.

In addition to shopping platforms, the dataset also includes other types of foreign pages, such as social media (Facebook, Instagram, etc.), online education platforms (Coursera, Udemy), news portals (BBC, CNN), and cloud service pages (AWS, Google Cloud). These pages have different design styles and functional requirements, and the focus of color application is also different. For example, social media pages are usually based on fresh and soft colors, online education platforms pay more attention to color brightness and reading comfort, and news portals tend to be concise and efficient in conveying information. These diverse page designs further enhance the representativeness of the dataset.

In order to ensure the diversity and quality of the data, the dataset also includes high-visit websites specific to certain countries or regions, such as Booking.com, TripAdvisor and other travel-related platforms. The design of these pages usually combines regional cultural characteristics and shows a color matching style different from the mainstream design. By integrating the above data sources, the dataset not only covers interface design in various fields but also can extract color features with unique meanings from different cultural backgrounds, providing a solid data foundation for building a color quality quantification model. The data example is shown in Figure 2.

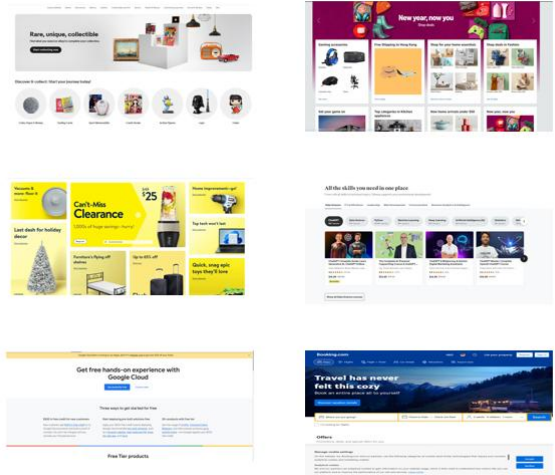

Figure 2 Dataset Example

*B. Experimental Results*

In interface design, the strong contrast relationship of hue plays an important role in the user's visual perception and attention guidance. In order to accurately quantify the angle range of strong contrast colors, this paper first experiments with different hue combinations as the research object, extracts their relative angle relationship on the color wheel, and analyzes the impact of hue contrast on visual impact based on experimental data. By statistically analyzing and fitting the changing trend of hue contrast intensity, a hue strong contrast angle distribution model suitable for interface design is constructed. The experimental results are shown in Figure 3.

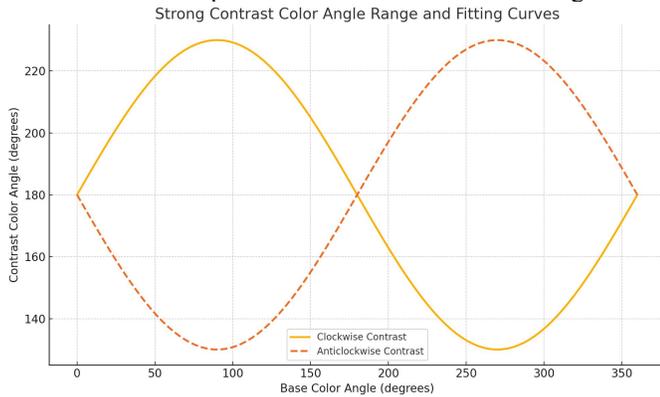

Figure 3 Strong Contrast Color Angle Range and Fitting Curves

Experimental results reveal that strong contrast angles exhibit a periodic and symmetrical pattern on the color wheel. In both clockwise and counterclockwise directions, angles close to 180° produce the most pronounced contrast, while deviation from 180° attenuates contrast intensity. This symmetry underscores the closed-loop nature of the color wheel and provides practical guidance for human–computer interaction (HCI) design. For instance, designers can enhance user focus by selecting color pairs near 180° for critical interface elements, while opting for lower contrast angles to reduce visual fatigue. Additionally, this study employs a convolutional neural network (CNN) to extract balance, continuity, integrity, and unity features from both the original images and CNN-processed outputs, as shown in Figure 4.

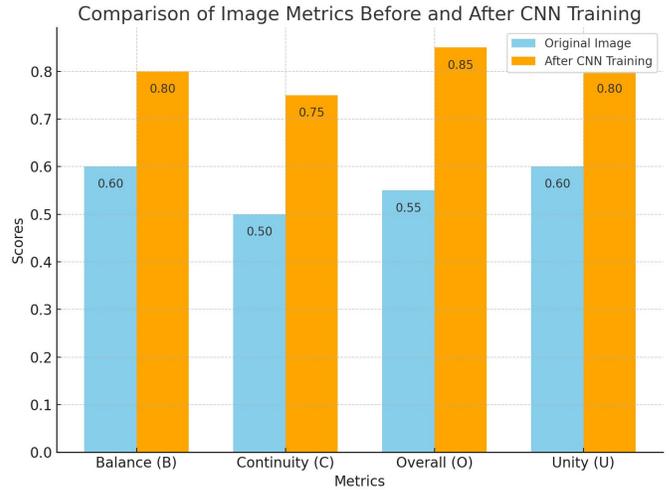

Figure 4 Comparison of Image Metrics Before and After CNN Training

From the experimental results, it can be seen that CNN-processed images yield marked improvements in four key indicators: balance, continuity, integrity, and unity. Balance rose from 0.60 to 0.80, indicating more symmetric designs that enhance visual comfort. Continuity increased from 0.50 to 0.75, reflecting reduced fragmentation and stronger coherence—critical for maintaining logical flow in dynamic interfaces. Integrity and unity similarly improved from 0.55 and 0.60 to 0.85 and 0.80, respectively, suggesting optimized layouts and consistent color schemes. These advances enhance both the scientific rigor of design and the overall aesthetic harmony of user interfaces.

Additionally, a questionnaire study was conducted in which both the CNN model and a group of 500 users evaluated the same interfaces. The findings of this comparative assessment are summarized in Table 1.

Table 1 Consistency analysis between CNN evaluation and user reviews

| Model | Average user rating | CNN average rating | Pearson | MSE | MAE |
| --- | --- | --- | --- | --- | --- |
| Contrast | 0.74 | 0.76 | 0.94 | 0.003 | 0.019 |
| Clarity | 0.72 | 0.75 | 0.93 | 0.002 | 0.018 |
| Color coordination | 0.70 | 0.73 | 0.96 | 0.001 | 0.014 |
| Visual appeal | 0.73 | 0.77 | 0.95 | 0.002 | 0.016 |

From The experimental data reveal a strong alignment between CNN evaluations and user ratings across all metrics. For Contrast, the average user rating (0.74) closely matches the CNN score (0.76), yielding a correlation coefficient of 0.94. Notably, its mean square error (MSE) is 0.003 and mean absolute error (MAE) is 0.019, demonstrating the model's precision. In Clarity and Color Coordination, CNN scores are similarly aligned with user perceptions, with correlation coefficients of 0.93 and 0.96, respectively. Color Coordination shows particularly low MSE (0.001) and MAE (0.014), indicating CNN's robust color analysis capabilities. For Visual Appeal, CNN's average score of 0.77 is slightly above the user average of 0.73, yet still highly correlated (0.95). Overall, these results confirm that CNN not only mirrors users' subjective ratings effectively but also delivers stable, consistent evaluations—making it an apt solution for scientifically assessing interface design quality.

## V. CONCLUSION

This paper proposes a quantitative evaluation model for the color quality of human-computer interaction interfaces by combining deep convolutional neural networks (CNNs), and comprehensively analyzes the impact of core features such as hue, brightness, and purity on user-perceived quality. Experimental results show that the CNN model can efficiently capture multi-level features in interface design and is highly consistent with user subjective ratings, especially in terms of contrast, clarity, color coordination, and visual appeal, providing a scientific and systematic solution for interface design quality evaluation.

The study also found that CNN has significant efficiency and accuracy advantages in evaluating complex interface designs. The robustness and generalization ability of the model were verified through experimental analysis of real interface samples from multiple industries. This data-driven quantitative method not only avoids the subjective problems in traditional evaluation methods but also provides designers with a quantifiable optimization basis, thereby improving the overall aesthetics of interface design and user experience quality.

However, this study still has certain limitations when dealing with dynamic interfaces or multimedia elements, especially the impact of user cultural background and personalized needs is not fully covered. In the future, multimodal learning methods can be further introduced to combine images, text, and user interaction data to establish a more comprehensive evaluation model. In addition, it is necessary to explore the application of the model in real-time scenarios, such as dynamically adjusting the interface color or layout to adapt to the user's immediate feedback.

Looking forward to the future, the interface quality evaluation method based on deep learning will be widely used in many fields such as smart homes, virtual reality, education platforms and medical systems. With the development of deep learning technology, the accuracy and adaptability of the evaluation model will be further improved, providing users with a more intelligent and personalized interactive experience, while injecting new impetus into the scientific and data-based development of human-computer interaction interface design.